\documentclass{Interspeech}

\interspeechcameraready

\usepackage{amssymb}
\usepackage{makecell}
\usepackage{multirow}
\usepackage{pifont}
\usepackage{cite}
\usepackage{tikz}
\newcommand{\cmark}{\ding{51}}
\newcommand{\xmark}{\ding{55}}
\let\OLDthebibliography\thebibliography
\renewcommand\thebibliography[1]{
  \OLDthebibliography{#1}
  \scriptsize 
  \setlength{\parskip}{1pt}
  \setlength{\itemsep}{0pt plus 0.05ex}
}
\usepackage{hyperref}

\definecolor{nodefill}{RGB}{213, 232, 212}
\definecolor{nodedraw}{RGB}{130, 179, 2}
\newcommand*\circled[1]{\tikz[baseline=(char.base)]{
            \node[fill=nodefill,shape=circle,draw=nodedraw,inner sep=0.3pt] (char) {#1};}
            }

\makeatletter
\renewcommand{\section}{\@startsection
  {section}%
  {1}%
  {}%
  {-0.5\baselineskip} 
  {0.25\baselineskip} 
  {}}
  
\renewcommand{\subsection}{\@startsection
  {subsection}%
  {2}%
  {}%
  {-0.5\baselineskip} 
  {0.2\baselineskip} 
  {}}

\renewcommand{\subsubsection}{\@startsection
  {subsubsection}%
  {3}%
  {}%
  {-0.5\baselineskip}%
  {0.2\baselineskip}%
  {}}%
\makeatother

\title{Personalized Fine-Tuning with Controllable Synthetic Speech from LLM-Generated Transcripts for Dysarthric Speech Recognition}

\author[affiliation={1}]{Dominik}{Wagner}
\author[affiliation={1}]{Ilja}{Baumann}
\author[affiliation={1}]{Natalie}{Engert}
\author[affiliation={2}]{Seanie}{Lee}
\author[affiliation={3}]{Elmar}{Nöth}
\author[affiliation={1}]{Korbinian}{Riedhammer}
\author[affiliation={1}]{Tobias}{Bocklet}

\affiliation{}{Technische Hochschule Nürnberg Georg Simon Ohm}{Germany}
\affiliation{}{Korea Advanced Institute of Science \& Technology}{Republic of Korea}
\affiliation{}{Friedrich-Alexander-Universität Erlangen-Nürnberg}{Germany}
\email{firstname.lastname@th-nuernberg.de, lsnfamily02@kaist.ac.kr, elmar.noeth@fau.de}

\usepackage{comment}

\begin{document}

\maketitle

\looseness=-1
\begin{abstract}
In this work, we present our submission to the Speech Accessibility Project challenge for dysarthric speech recognition. 
We integrate parameter-efficient fine-tuning with latent audio representations to improve an encoder-decoder ASR system. 
Synthetic training data is generated by fine-tuning Parler-TTS to mimic dysarthric speech, using LLM-generated prompts for corpus-consistent target transcripts. 
Personalization with x-vectors consistently reduces word error rates (WERs) over non-personalized fine-tuning. 
AdaLoRA adapters outperform full fine-tuning and standard low-rank adaptation, achieving relative WER reductions of $\sim$23\% and $\sim$22\%, respectively. 
Further improvements ($\sim$5\% WER reduction) come from incorporating wav2vec 2.0-based audio representations. 
Training with synthetic dysarthric speech yields up to $\sim$7\% relative WER improvement over personalized fine-tuning alone.
\end{abstract}

\section{Introduction}
Dysarthria is a motor speech disorder characterized by impaired control of the muscles involved in speech production, which can result from damage to the central or peripheral nervous system. 
This impairment leads to difficulties with articulation, phonation, respiration, resonance, and prosody \cite{darley1969dysarthria}. 
Dysarthria manifests differently depending on the underlying neurological condition. 
For instance, hypokinetic dysarthria, commonly associated with Parkinson’s disease (PD), is characterized by monotony of pitch and loudness, reduced stress, imprecise consonants, a breathy or harsh voice, and rapid bursts of speech \cite{darley1969dysarthria}. 
PD is a chronic and progressive neurodegenerative disorder that primarily affects the motor system due to the loss of dopamine-producing neurons in the substantia nigra region of the brain \cite{lang1998parkinson,fearnley1991parkinsons}. 
About 70\% of PD patients develop speech impairments, which tend to worsen over time \cite{hartelius2009parkinson,klawans1986parkinson}.
Dysarthria is also associated with other neurological conditions, such as amyotrophic lateral sclerosis (ALS) and cerebellar disorders like cerebral palsy. 
These disorders result in distinct patterns of speech dysfunction, such as mixed dysarthria in ALS and ataxic dysarthria in cerebellar disorders \cite{darley1969dysarthria}.

Despite advancements in automatic speech recognition (ASR) systems, they remain ill-suited for dysarthric speech due to the high variability of speech patterns across individuals and the limited availability of large, high-quality datasets \cite{moore18_interspeech,xiong2019dysarthric,leung24_interspeech,christensen13b_interspeech}.  
Conventional ASR systems, optimized for typical speech, struggle to generalize to atypical speakers, resulting in poor recognition accuracy. 
This limitation is particularly detrimental for individuals with dysarthria, as speech often serves as a critical mode of interaction with devices like smartphones and smart home systems \cite{de_russis_impact_2019}.

To address data scarcity, various data augmentation techniques have been proposed for training ASR systems for dysarthric speech. 
Audio-level augmentations, such as speed, pitch, tempo, volume, and vocal tract length perturbations, have proven effective in increasing training data diversity and robustness \cite{vachhani18_interspeech,geng20_interspeech,bhat22_interspeech,xiong2019phonetic}. 
SpecAugment \cite{park19e_interspeech} and dynamic augmentation techniques inspired by it have also been shown to enhance ASR accuracy for dysarthric speech \cite{wang23qa_interspeech,leung24_interspeech,bhat22_interspeech}. 
Domain-specific methods, such as Dysarthric SpecAugment, simulate impaired speech characteristics from typical speech using domain knowledge \cite{bhat22_interspeech}. 
Text-to-speech (TTS) and voice conversion systems have also been employed to mitigate data scarcity by generating synthetic dysarthric speech \cite{leung24_interspeech,wang23qa_interspeech,soleymanpour2022synthesizing,jin2023adversarial,harvill2021synthesis}.
 

Speaker adaptation is a widely used method for personalizing ASR systems to dysarthric speakers. 
For example, in \cite{baskar22b_interspeech}, feature-space maximum likelihood linear regression \cite{GALES199875} and x-vectors \cite{snyder18xvector} are used as auxiliary features for self-supervised models like wav2vec 2.0 \cite{baevski20w2v2} to capture speaker-specific characteristics.  
Other approaches include fine-tuning speaker-independent models on individual speakers \cite{takashima2020adaptation,tomanek2023analysis,shor19_interspeech} and leveraging latent features \cite{geng23_interspeech,wang24x_interspeech,hernandez22_interspeech,hu2024ssl}.
More recently, speaker-independent systems \cite{singh2025robust} and multi-task approaches with auxiliary classification tasks \cite{zheng24c_interspeech} have been used. 

Recent research shows the importance of personalization, synthetic data generation, and augmentation techniques in improving ASR performance for dysarthric speech. 
Building on these findings, this paper explores a lightweight personalization approach for large end-to-end ASR systems developed for small academic lecture corpora \cite{wagner2024personal}, leveraging a modified Whisper \cite{radford2022whisper} architecture with speaker adaptation capabilities and parameter-efficient fine-tuning. 
Following previous works~\cite{wang23qa_interspeech,leung24_interspeech,soleymanpour2022synthesizing}, we integrate SpecAugment into the training pipeline and explore the effectiveness of supplementing the relatively small amount of training data with artificial examples generated by a text-to-speech (TTS) synthesis system. 
Specifically, we fine-tune 
Parler-TTS \cite{lyth2024parlertts}, an autoregressive transformer \cite{transformer} model operating on quantized audio units, to generate synthetic training data for dysarthric speech. 
In contrast to previous works \cite{soleymanpour2022synthesizing,leung24_interspeech}, Parler-TTS allows fine-grained control over attributes such as gender, noise level, and speaking rate through text descriptions, enabling the creation of diverse dysarthric speech datasets. 
Furthermore, Parler-TTS can be effectively fine-tuned using attributes such as pitch and speech quality derived from dysarthric speech data in combination with novel target transcripts generated by large language models (LLMs). 
Our experiments show that personalization with x-vectors consistently improves performance compared to non-personalized fine-tuning. 
AdaLoRA outperforms other training methods and further improvements can be achieved by incorporating additional audio representations into the ASR system. 
We observe that the effectiveness of synthetic data varies across etiologies, with performance gains depending on both the specific condition and the TTS model used for generation. 
\section{Data}
We employ the Speech Accessibility Project \cite{hasegawa2024sap} (SAP) dataset for all experiments, with models submitted to the associated challenge.\footnote{\url{https://eval.ai/web/challenges/challenge-page/2362/overview}}
Specifically, we use the SAP \texttt{2024-11-30} data package, a dataset curated for the development of dysarthric ASR systems, which includes English dysarthric speech from individuals with PD, ALS, cerebral palsy, down syndrome, and stroke. 
The \texttt{2024-11-30} version of the data contains speech recordings from a total of 430 participants, with 374 assigned to the training set and 56 to the development (\textit{dev}) set. 
Each participant recorded an average of 350 utterances with an average duration of $7.9\pm9.0$ seconds, drawn from virtual assistant commands, sentences from novels, and spontaneous speech. 
The combined duration of the recordings is $\sim$333 hours with $\sim$290 hours allocated to the training set and $\sim$43 hours allocated to the \textit{dev} set. 
The majority of the data ($\sim$74\%) are recordings from participants with PD. 
The remainder is distributed between ALS ($\sim$14\%), cerebral palsy ($\sim$6\%), down syndrome ($\sim$4\%), and stroke ($\sim$2\%).  

Most experimental results reported in our work are based on the \textit{dev} set. 
We also report some results on the unseen \textit{test1} set as part of the SAP challenge. 
However, due to upload restrictions and time constraints, we were unable to generate scores for every experiment.
\section{Method}
\begin{figure}[t]
    \centering
    \includegraphics[width=0.99\linewidth]{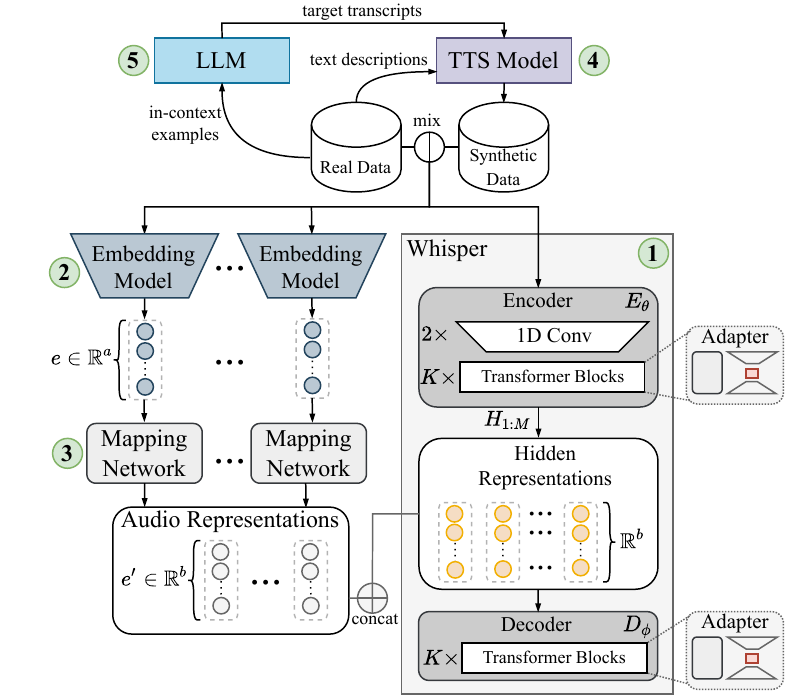}
    \vspace{-3mm}
    \caption{Overview of the proposed ASR system combining the incorporation of audio representations, fine-tuning, and synthetic training data generation.}
    \label{fig:overview}
    \vspace{-6mm}
\end{figure}

Figure~\ref{fig:overview} provides an overview of the components used in this work. 
The speech recognition system \circled{\textbf{1}} is based on the approach introduced in \cite{wagner2024personal}. 
It employs Whisper \cite{radford2022whisper} as the ASR backbone and a personalization component based on a small neural network \circled{\textbf{3}}, which projects audio representations obtained from a pre-trained and frozen embedding model (e.g. x-vectors \cite{snyder18xvector}) \circled{\textbf{2}} into the latent space of the decoder. 
The model is trained to recognize dysarthric speech using either full fine-tuning (FFT) or parameter-efficient fine-tuning with LoRA \cite{hu2022lora} and AdaLoRA \cite{zhang2023adalora}. 
Synthetic speech data is generated with a fine-tuned Parler-TTS model \cite{lyth2024parlertts} \circled{\textbf{4}} using LLMs \circled{\textbf{5}} to generate new target transcripts that closely align with the real data. 
\vspace{-0.15in}
\subsection{Model adaptation}
Whisper employs a transformer-based encoder-decoder architecture~\cite{transformer}. 
Given log-Mel spectrogram features $X_{1:T} = (x_1, \dots, x_T)$, the encoder maps them to latent representations:
\vspace{-2mm}
\begin{equation}
    H_{1:M} = E_{\theta}(X_{1:T}), \quad H_{1:M} \in \mathbb{R}^{M \times b},
\end{equation}
where $E_{\theta}$ is the encoder with parameters $\theta$, and $M = T/2$ after subsampling. 
The decoder then estimates a probability distribution for the next token $y_i$ over the vocabulary, given $H_{1:M}$ and the previous tokens $y_{<i}=(y_1, \ldots, y_{i-1})$:
\vspace{-2mm}
\begin{equation}
    P(y_i | y_{<i}, H_{1:M}) = D_{\phi}(y_{<i}, H_{1:M}), \text{ for } i\in[T^\prime],
\end{equation}
where $D_{\phi}$ is the decoder with parameters $\phi$ and $[T^\prime]=\{1,\ldots, T^\prime\}$. 
The architecture is modified by incorporating audio representations. 
Given input features $Z_{1:L} = (z_1, \dots, z_L)$, multiple audio embedding models  map them to a set of fixed-dimensional latent representations. 
Each embedding model $f_{\psi_i}$ generates a sequence of $U_i$ latent vectors. The sequence is averaged to obtain a single fixed-dimensional vector:
\vspace{-2mm}
\begin{equation}
e^{(i)} =
\frac{1}{U_i} \sum_{u=1}^{U_i} f_{\psi_i}(Z_{1:L})_u \in \mathbb{R}^{a_i}, \text{ for } i \in [N],
\end{equation}
where $f_{\psi_i}(Z_{1:})_u\in\mathbb{R}^{a_i}$ is $u$-th row of $f_{\psi_i}(Z_{1:L})\in\mathbb{R}^{U_i \times a_i}$.
Each audio representation $e^{(i)}$ is then projected into the decoder's latent space via a dedicated mapping network:
\vspace{-2mm}
\begin{equation}
    e'^{(i)} = W_2^{(i)} \sigma_{\tau}(W_1^{(i)} e^{(i)} + b_1^{(i)}) + b_2^{(i)}, \quad e'^{(i)} \in \mathbb{R}^{b},
\end{equation}
where \(\{ (W_j^{(i)},  b_j^{(i)})\}_{j=1}^2 \) are trainable parameters specific to each mapping network, and \( \sigma_{\tau} \) is Tanh. 
Finally, all the representations $\{e'^{(i)}\}_{i=1}^{N}$ are concatenated with the decoder’s input:
\vspace{-2mm}
\begin{equation}
    \tilde{H}_{1:N+M} = \text{concat}(e'^{(1)}, \dots, e'^{(N)}, H_{1:M}).
\end{equation}
During training, only the adapter modules and mapping networks are updated, all other components remain frozen. 
\subsection{Synthetic dysarthric speech generation}
We fine-tune the \texttt{parler-tts-mini-v1} version of Parler-TTS \cite{lacombe2024parlertts}, a 880M parameter TTS model trained on $\sim$45k hours of audio data. 
Parler-TTS is an open-source and open-weight implementation of \cite{lyth2024parlertts}, which is based on the MusicGen \cite{copet2023musicgen} architecture. 
The model has three core stages: First, a pre-trained and frozen text encoder based on FLAN-T5 \cite{chung2022flant5} converts a text description of the target audio into a sequence of hidden representations. 
Second, a transformer language model acts as a decoder that autoregressively generates audio tokens. 
The model relies entirely on the text description for gender category, tone, and speech patterns. 
Cross-attention is used to condition the model on these text descriptions.
The text prompts representing the spoken utterance are pre-pended to the decoder inputs. 
Finally, a neural audio codec is used to synthesize waveforms from the generated audio tokens. 
The open implementation of Parler-TTS employs the Descript Audio Codec (DAC) \cite{kumar2023highfidelity} instead of EnCodec \cite{defossez2023high} for audio synthesis. 
\subsubsection{Target prompt generation}
We use both Phi-3 \cite{abdin2024phi3report} (\texttt{Phi-3-medium-4k-instruct}) and Llama 3.1 70B \cite{grattafiori2024llama3report} (\texttt{Llama-3.1-70B-Instruct}) for generating English target transcripts. 
To seed the generation of diverse synthetic data that roughly follows the topic and phrasing distribution of the SAP dataset, we follow \cite{li2023phi15report} and instruct the LLM to generate new sentences based on ten randomly sampled utterances from the SAP training set. 
The goal is to increase data diversity in a controlled manner by generating sentences that closely align with the SAP dataset in terms of vocabulary, topics, and phrasing, while ensuring they are not identical.
The SAP dataset does not include a gender attribute. 
However, we aim to make gender a controllable feature in Parler-TTS since the model has been pre-trained with binary gender information. 
To address this, we employ a pre-trained classifier to predict gender labels for the SAP data. 
The model is based on an ECAPA-TDNN \cite{desplanques20_interspeech} speaker verification model that was fine-tuned to predict binary gender labels using the VoxCeleb2 development set.
\section{Experiments and results}
Our experiments were evaluated using the SAP challenge evaluation scheme, which is based on two metrics: WER and semantic score (SemScore). 
SemScore is a weighted sum of BERTscore \cite{zhang2020bertscore}, phonetic distance, and natural language inference probability \cite{chen2023menli}.
For each experiment, we explored multiple hyperparameter configurations. 
The models trained with LoRA used learning rates $\eta \in \{10^{-3}, 10^{-4}, 2 \cdot 10^{-5}\} $, scaling parameters $\alpha \in \{32, 64, 128\}$, and ranks $r \in \{ 8, 16, 32\}$. 
Since AdaLoRA showed more stable behavior, we set $\eta = 10^{-3}$ and $\alpha = 32$ and explored rank configurations of $r_{\text{initial}} \in \{ 12, 16, 24, 32, 64 \}$ and $r_{\text{target}} \in \{ 8, 12, 16, 24, 32 \}$. 
FFT was conducted using the learning rates $\eta \in \{10^{-4}, 5 \cdot 10^{-5}, 5 \cdot 10^{-6}\}$.  
We report the results under the best hyperparameter configuration on the \textit{dev} set for each experiment in the subsequent tables.  
All experiments were conducted using the largest available Whisper model (\texttt{whisper-large-v3}) as the ASR backbone. 
Adapter modules were attached to the query and value matrices of each transformer block in the encoder and decoder. 
Speaker characteristics were incorporated via x-vectors extracted from a model trained on VoxCeleb 1 and 2 \cite{nagrani17_interspeech,chung18b_interspeech}. 
All experiments used a dropout probability of 0.1 in the adapter modules and mapping networks. 
The models were trained for 15 epochs. 
The best model was chosen based on the lowest WER on a randomly sampled subset ($\sim$10\%) of the \textit{dev} set. 
Performance was evaluated every 2000 training steps. 
We used the AdamW optimizer ($\lambda = 10^{-4}$, $\epsilon = 10^{-8}$, $\beta_1 = 0.99$, $\beta_2 = 0.999$) with a linear rate schedule, a warm-up phase of 500 steps, and initial learning rates depending on the training method. 
Greedy decoding was used in all experiments. 

We fine-tuned Parler-TTS for 10 epochs on the SAP training data using the AdamW optimizer ($\lambda = 10^{-2}$, $\epsilon = 10^{-8}$, $\beta_1 = 0.9$, $\beta_2 = 0.99$) with a constant learning rate of $9.5 \cdot 10^{-4}$ and a warm-up phase of 2k steps. 
To generate synthetic dysarthric speech data, attributes characterizing the speaker were randomly sampled from the SAP training set and paired with LLM-generated prompts. Parler-TTS occasionally failed to produce intelligible speech for certain attribute-prompt combinations. To address these cases, we transcribed all synthetic utterances using Whisper (\texttt{whisper-medium}) and discarded those with a WER of 35 or higher.
\vspace{-0.05in}
\subsection{Main results}

\begin{table}[t]
\vspace{-3.5mm}
\centering
\caption{Comparison of training methods, personalization and SpecAugment.  Results in parentheses indicate the non-public \textit{test1} set, all other results were computed on the \textit{dev} set. 
}
\vspace{-3mm}
\label{tab:sapc_main}
\setlength{\tabcolsep}{7pt}
\resizebox{0.48\textwidth}{!}{
\begin{tabular}{ccccccc}
\toprule
\multirow{2}{*}{\textbf{\#}} & 
\multirow{2}{*}{\makecell{\textbf{Training} \\ \textbf{Method}}} & 
\multirow{2}{*}{\makecell{\textbf{Personal-} \\ \textbf{ization}}} & 
\multirow{2}{*}{\makecell{\textbf{Spec-} \\ \textbf{Augment}}} & 
\multirow{2}{*}{\makecell{\textbf{WER} \\ ($\downarrow$)}} & 
\multirow{2}{*}{\makecell{\textbf{SemScore} \\ ($\uparrow$)}}  \\
 &  &  &  &  & \\
\midrule
1  & None & \xmark & \xmark & 12.31 & 84.63 \\
\midrule
2  & LoRA & \xmark & \xmark & 10.87 & 88.92 \\
3  & LoRA & \xmark & \cmark & 13.28 & 88.47 \\
4  & AdaLoRA & \xmark & \xmark & 11.73 & 87.02 \\
5  & AdaLoRA & \xmark & \cmark & 11.32 & 87.32 \\
6  & FFT & \xmark & \xmark & 11.77 & 87.26 \\
7  & FFT & \xmark & \cmark & 10.57 & 88.13 \\
\midrule
8  & LoRA & \cmark & \xmark & 10.52 & 88.89 \\
9  & LoRA & \cmark & \cmark & 11.09 & 88.93 \\
10 & \makecell{AdaLoRA} & \cmark & \xmark & \makecell{\textbf{8.05} \scriptsize{\textit{(10.65)}}} & \makecell{91.29 \scriptsize{\textit{(88.33)}}} \\
11 & \makecell{AdaLoRA} & \cmark & \cmark & \makecell{8.12 \scriptsize{\textit{(11.36)}}} & \makecell{\textbf{92.31} \scriptsize{\textit{(88.93)}}} \\
12 & FFT & \cmark & \xmark & 11.54 & 86.69 \\
13 & FFT & \cmark & \cmark & 10.27 & 88.50 \\
\bottomrule
\end{tabular}
}
\vspace{-6mm}
\end{table}

The goal of the experiments in Table~\ref{tab:sapc_main} is evaluating the impact of personalization and SpecAugment, while also comparing the performance of LoRA, AdaLoRA, and FFT. 
The first row (\#1) shows the results when inference on the underlying base model (\texttt{whisper-large-v3}) is performed without any fine-tuning. 
Applying SpecAugment during training leads to WER and SemScore improvements with AdaLoRA and FFT in the non-personalized setting (\#2-7).
However, for LoRA without personalization, it increases WER from 10.87 to 13.28. 
In the personalized setting (\#8-13), AdaLoRA shows a slight degradation in WER from 8.05 to 8.12 but an improvement in SemScore from 91.29 to 92.31. 
Without personalization, FFT yields the lowest WER (10.57), while LoRA and AdaLoRA perform worse. 
However, in the personalized setting, AdaLoRA outperforms both LoRA and FFT, achieving the lowest WER (8.05 in \#10) and highest SemScore (92.31 in \#11). 

The results of experiment \#10 represent a $\sim$31\% reduction in WER and a $\sim$6\% increase in SemScore compared to the corresponding experiment without personalization (\#4). 
On the non-public \textit{test1} set, WER deteriorated more significantly with SpecAugment (\#11) compared to the experiment without it (\#10), while SemScore remained higher in \#11. 
Even though, SpecAugment does not yield consistent performance gains, we found that it somewhat stabilizes training with respect to hyperparameter choices, particularly the initial and target ranks in AdaLoRA fine-tuning. 
\vspace{-0.05in}
\subsection{Generalization beyond Whisper}
To evaluate the generalization ability of x-vector-based personalization beyond Whisper, we applied it to HuBERT \cite{weining21hubert} as well. 
Specifically, we used the largest available HuBERT model (\texttt{hubert-xlarge-ll60k}; 1B parameters) as the base encoder, followed by a bidirectional long short-term memory (BiLSTM) \cite{graves2005bilstm} block and a connectionist temporal classification (CTC) objective \cite{graves2006ctc}. 
Personalization was integrated by concatenating the outputs of the HuBERT model with speaker vectors, before being fed into the BiLSTM layer. 
Without personalization, the model achieved a WER of 9.31 and a SemScore of 91.06 on the \textit{dev} set. 
When incorporating personalization, the model improved, achieving a WER of 8.79 and a SemScore of 91.42 on the \textit{dev} set. 
\vspace{-0.05in}
\subsection{Adding additional audio representations}
\begin{table}[ht]
\vspace{-3mm}
\centering
\caption{Results with wav2vec 2.0 audio representations. All experiments were conducted using AdaLoRA w/o SpecAugment.}
\vspace{-3mm}
\label{tab:sapc_audio_emb}
\setlength{\tabcolsep}{8pt}
\resizebox{0.48\textwidth}{!}{
\begin{tabular}{cccccc}
\toprule
\multirow{2}{*}{\textbf{\#}} &
\multirow{2}{*}{\makecell{\textbf{Personal-} \\ \textbf{ization}}} & 
\multirow{2}{*}{\makecell{\textbf{Audio} \\ \textbf{Representation}}} & 
\multirow{2}{*}{\makecell{\textbf{WER} \\ ($\downarrow$)}} & 
\multirow{2}{*}{\makecell{\textbf{SemScore} \\ ($\uparrow$)}} \\
 &  &  &  &  \\
\midrule
1.1 & \cmark & base (layer 2) & 7.80 & 91.16 \\
1.2 & \cmark & base (layer 6) & 7.86 & 91.31 \\
1.3 & \cmark & base (layer 12) & \textbf{7.68} & \textbf{91.38} \\
1.4 & \cmark & etiology & \makecell{8.07 \scriptsize{\textit{(10.78)}}} & \makecell{91.17 \scriptsize{\textit{(88.21)}}} \\
1.5 & \cmark & category & \makecell{8.09 \scriptsize{\textit{(10.67)}}} & \makecell{91.27 \scriptsize{\textit{(88.35)}}}\\
\midrule
1.6 & \xmark & base (layer 2) & 9.30 & 89.36 \\
1.7 & \xmark & base (layer 6) & 9.53 & 88.78 \\
1.8 & \xmark & base (layer 12) & 8.09 & 91.21 \\
1.9 & \xmark & etiology & 11.55 & 87.13 \\
1.10 & \xmark & category & 8.83 & 90.70 \\
\bottomrule
\end{tabular}
}
\vspace{-6mm}
\end{table}

Since incorporating speaker vectors during training improved performance (cf. Table~\ref{tab:sapc_main}), a natural extension is to explore the effect of additional audio representations. 
Inspired by \cite{gimenogomez2024unveiling,wagner24b_interspeech}, we trained linear classifiers on top of a pre-trained wav2vec 2.0 model (\texttt{wav2vec2-base}; 94M parameters) to predict metadata from the SAP dataset. 
Specifically, we trained classifiers to infer etiology labels and utterance categories.  
We then extracted latent representations at the last transformer layer of the underlying wav2vec 2.0 backbone and used them as additional input features to the ASR system. 
Additionally, we extracted representations from the non-fine-tuned wav2vec 2.0 base model at layers 2, 6, and 12. 

We extended experiment \#10 from Table~\ref{tab:sapc_main} (AdaLoRA with personalization, without SpecAugment) due to the disproportionate WER increase on the \textit{test1} set observed in experiment \#11. 
Table~\ref{tab:sapc_audio_emb} presents the results of integrating additional audio representations. 
In the upper half (\#1.1-1.5), where personalization is applied, several configurations outperform experiment \#10 from Table~\ref{tab:sapc_main}. 
However, using representations fine-tuned for etiology and category classification embeddings did not yield improvements (\#1.4 and \#1.5).  
The best-performing experiment is AdaLoRA with personalization and audio representations extracted at layer 12 of the base model, achieving 7.68 WER and 91.38 SemScore. 
These results indicate that lower-layer representations (e.g., layer 2 and layer 6) are suboptimal in this context, achieving slightly higher WERs and lower SemScores. 
This aligns with previous findings that deeper layers in wav2vec 2.0 encode more phonetic and word content \cite{pasad2021layer}, which may be more beneficial for ASR.
The lower half of the table, where personalization is is not used, shows that audio embeddings alone do not improve performance. 
However, when combined with personalization, synergies emerge, leading to overall performance gains. 
\vspace{-0.05in}
\subsection{Adding synthetic speech data}
The effect of supplementing the training data with increasing amounts of synthetic speech are shown in Table~\ref{tab:sapc_synth_all}.  
Following \cite{leung24_interspeech}, we report the results for ratios between 10\% and 100\% of additional synthesized training data. 
In the setup without SpecAugment (\#2.1-2.3), adding Parler-TTS-generated synthetic speech from LLM-generated target prompts improves performance over experiment \#10 from Table~\ref{tab:sapc_main}. 
At 10\% synthetic data, WER drops to 7.71, and SemScore increases to 91.58. 
Increasing the synthetic ratio to 50\% further reduces WER to 7.57 and boosts SemScore to 92.25. 
At 100\%, the best result is achieved, with a WER of 7.47 and SemScore of 92.39, both improvements over experiment \#10 in Table~\ref{tab:sapc_main}. 
Experiments \#2.4-2.6 show that SpecAugment does not improve performance in this setting. 
However, WERs also decrease with increasing amounts of synthetic data.  
Experiments \#2.7-2.9 use transcripts from the SAP training set as prompts for the TTS system instead of LLM-generated prompts. 
These experiments show worse performance compared to speech synthesis based on the more diverse LLM-generated prompts (\#2.1-2.3). 
Experiments \#2.10-2.12 follow experiment \#1.3 from Table~\ref{tab:sapc_audio_emb}, incorporating additional audio representations obtained from the last layer of a pre-trained wav2vec 2.0 system. 
While effective without synthetic data, results did not improve when synthetic data was used. 

\begin{table}[t]
\vspace{-0.1in}
\centering
\caption{Results with varying portions of synthetic training data using AdaLoRA fine-tuning. All experiments employ x-vector based personalization.}
\vspace{-3mm}
\label{tab:sapc_synth_all}
\setlength{\tabcolsep}{1.5pt}
\resizebox{0.98\columnwidth}{!}{
\begin{tabular}{cccccccc}
\toprule
\multirow{2}{*}{\textbf{\#}} &
\multirow{2}{*}{\makecell{\textbf{TTS} \\ \textbf{Model}}} &
\multirow{2}{*}{\makecell{\textbf{LLM} \\ \textbf{Prompt}}} &
\multirow{2}{*}{\makecell{\textbf{Spec-} \\ \textbf{augment}}} &
\multirow{2}{*}{\makecell{\textbf{Audio} \\ \textbf{Repr.}}} &
\multirow{2}{*}{\makecell{\textbf{Synth} \\ \textbf{Data}}} &
\multirow{2}{*}{\makecell{\textbf{WER} \\ ($\downarrow$)}} &
\multirow{2}{*}{\makecell{\textbf{SemScore} \\ ($\uparrow$)}} \\
 &  &  &  &  &  &  \\
\midrule
2.1 & Parler-TTS & \cmark & \xmark & \xmark & 10\% & 7.71 & 91.58 \\
2.2 & Parler-TTS & \cmark & \xmark & \xmark & 50\% & 7.57 & 92.25 \\
2.3 & Parler-TTS & \cmark & \xmark & \xmark & 100\% & \textbf{7.47} & \textbf{92.39} \\
\midrule
2.4 & \makecell{Parler-TTS}& \cmark & \cmark & \xmark & 10\% & \makecell{9.28 \scriptsize{\textit{(11.68)}}} & \makecell{90.65 \scriptsize{\textit{(87.32)}}} \\
2.5 & \makecell{Parler-TTS} & \cmark & \cmark & \xmark & 50\% & \makecell{8.49 \scriptsize{\textit{(11.68)}}} & \makecell{91.65 \scriptsize{\textit{(88.62)}}} \\
2.6 & Parler-TTS & \cmark & \cmark & \xmark & 100\% & 7.92 & 91.17 \\
\midrule
2.7 & Parler-TTS & \xmark & \xmark & \xmark & 10\% & 8.44 & 91.83 \\
2.8 & Parler-TTS & \xmark & \xmark & \xmark & 50\% & 7.84 & 91.50 \\
2.9 & Parler-TTS & \xmark & \xmark & \xmark & 100\%& 8.43 & 90.03 \\
\midrule
2.10 & Parler-TTS & \cmark & \xmark & \cmark & 10\% & 8.17 & 90.66 \\
2.11 & Parler-TTS & \cmark & \xmark & \cmark & 50\% & 8.15 & 90.93 \\
2.12 & Parler-TTS & \cmark & \xmark & \cmark & 100\% & 8.27 & 91.79 \\
\midrule
2.13 & OpenVoice & \cmark & \xmark & \xmark & 10\% & 8.24 & 91.56 \\
2.14 & \makecell{OpenVoice} & \cmark & \xmark & \xmark & 50\% & \makecell{8.35 \scriptsize{\textit{(12.61)}} } & \makecell{90.48 \scriptsize{\textit{(86.13)}}} \\
2.15 & OpenVoice & \cmark & \xmark & \xmark & 100\% & 9.14 & 89.25 \\
\bottomrule
\end{tabular}
}
\vspace{-5mm}
\end{table}

In addition to Parler-TTS, we used OpenVoice \cite{qin2024openvoice} in a zero-shot setting by selecting reference speakers from randomly sampled utterances in the SAP training set to extract style information. 
The target transcripts were based on the same LLM-generated prompts used in Parler-TTS. 
However, in our experiments (\#2.13–2.15), OpenVoice-generated synthetic data resulted in higher WERs, with 8.24 at 10\% synthetic data compared to 7.71 for Parler-TTS. 
Performance further degrades as more synthetic data is added, with WER rising to 9.14 at 100\%.  

To further assess the impact of synthetic data augmentation, we compared the best results from Table~\ref{tab:sapc_synth_all} (\#2.1–2.3) to experiment \#10 from Table~\ref{tab:sapc_main} at the etiology level. Table~\ref{tab:sapc_synth_by_etiology} shows that the effectiveness of synthetic data varies by etiology, with the largest gains observed for ALS, while improvements for other conditions were smaller in relative terms. 
\begin{table}[ht]
\vspace{-2mm}
\centering
\caption{Comparison of results by etiology.}
\vspace{-3mm}
\label{tab:sapc_synth_by_etiology}
\setlength{\tabcolsep}{2pt}
\resizebox{0.98\columnwidth}{!}{
\begin{tabular}{c|cc|cc|cc|cc|cc}
\toprule
\multirow{2}{*}{\textbf{\#}} &  
\multicolumn{2}{c|}{\textbf{ALS}} &  
\multicolumn{2}{c|}{\textbf{Cerebral Palsy}} &  
\multicolumn{2}{c|}{\textbf{Down Syndrome}} &  
\multicolumn{2}{c|}{\textbf{Parkinson}} &  
\multicolumn{2}{c}{\textbf{Stroke}} \\  
& WER & SemScore & WER & SemScore & WER & SemScore & WER & SemScore & WER & SemScore \\  
\midrule
10 & 6.54 & 90.14 & 18.68 & 78.19 & 22.95 & 74.73 & 7.01 & 94.23 & 9.04 & 93.42 \\ 
\midrule
2.1 & 5.27 & 90.72 & 17.47 & 78.31 & 23.59 & 74.02 & 6.62 & 94.44 & 8.93 & 93.50 \\  
2.2 & 4.73 & 92.11 & 16.26 & 79.52 & 21.73 & 75.83 & 6.68 & 94.86 & 9.01 & 94.00 \\  
2.3 & 4.91 & 92.52 & 16.47 & 79.74 & 22.10 & 75.76 & 6.50 & 94.97 & 9.18 & 93.62 \\  
\bottomrule
\end{tabular}
}
\vspace{-3mm}
\end{table}

\subsection{Limitations}
Although, our approach improves dysarthric ASR through personalized fine-tuning and synthetic data augmentation, several limitations remain. 
First, the effectiveness of synthetic dysarthric speech varies by etiology, indicating that Parler-TTS and its textual description of voice attributes may not fully capture the complexity of real dysarthric speech patterns.  
Moreover, due to time and resource constraints, we were only able to obtain the best results with synthetic data after the challenge had already concluded. 
As a result, we were unable to score our best models on the unseen \textit{test1} set, leaving uncertainty about how well these models would generalize to this dataset.

\section{Conclusions}
This work explored improving dysarthric speech recognition through personalized fine-tuning with speaker vectors, synthetic speech augmentation, and the integration of additional audio representations. 
We fine-tuned Parler-TTS to generate controllable synthetic dysarthric speech using LLM-generated transcripts and incorporated personalization via x-vectors. 
Our results show that personalization with AdaLoRA significantly reduces WER, achieving up to a $\sim$31\% improvement over non-personalized fine-tuning. 
Additionally, integrating wav2vec 2.0-based audio representations further reduced WER by $\sim$5\%. 
Synthetic dysarthric speech contributed up to $\sim$7\% additional improvement on the overall dataset, but its effectiveness varied by etiology and TTS model.
\newpage
\bibliographystyle{IEEEtran}
\bibliography{references}
\end{document}